\documentclass[twocolumn,showpacs,paper,amsmath,amssymb]{revtex4}
\usepackage{epsfig}
\usepackage{dcolumn}

\usepackage{graphicx}
\usepackage{dcolumn}
\usepackage{bm}

\newcommand{\be}{\begin{equation}}
\newcommand{\ee}{\end{equation}}
\newcommand{\bea}{\begin{eqnarray}}
\newcommand{\eea}{\end{eqnarray}}

\newcommand{\integer}{\relax{\rm I\kern-.18em N}}

\begin{document}

\title{Gauge invariant counterparts \\and quantization of systems under holonomic constraints}

\author{M. I. Krivoruchenko$^{1,2}$, Amand Faessler$^{2}$, A. A. Raduta$^{2,3,4}$, C. Fuchs$^{2}$}

\affiliation{
$^{1}$Institute for Theoretical and Experimental Physics$\mathrm{,}$
B. Cheremushkinskaya 25$\mathrm{,}$ 117259 Moscow, Russia \\
$^{2}$Institut f\"{u}r Theoretische Physik$\mathrm{,}$ Universit\"{a}t T\"{u}bingen$\mathrm{,}$
Auf der Morgenstelle 14$\mathrm{,}$ D-72076 T\"{u}bingen$\mathrm{,}$ Germany \\
$^{3}$Department of Theoretical Physics and Mathematics$\mathrm{,}$ Bucharest University$\mathrm{,}$ 
Bucharest$\mathrm{,}$ POBox MG11$\mathrm{,}$ Romania \\
$^{4}$Institute of Physics and Nuclear Engineering$\mathrm{,}$ Bucharest$\mathrm{,}$ POBox MG6$\mathrm{,}$ Romania \\
}

\begin{abstract}
Systems under holonomic constraints are classified within the generalized Hamiltonian framework as second-class constraints systems. 
We show that each system of point particles with holonomic constraints has a hidden gauge symmetry which allows to quantize it 
in the original phase space as a first-class constraints system. The proposed method is illustrated with quantization of
a point particle moving on an $n-1$-dimensional sphere $S^{n-1}$ as well as its field theory analog the $O(n)$ nonlinear sigma model.
\end{abstract}


\maketitle

A systematic study of the generalized Hamiltonian dynamics was made by Dirac
\cite{DIRA} who classified the constraint structure and developed the corresponding 
operator quantization schemes of the constrained Hamiltonian systems. 

Gauge theories appear under first-class constraints. The Dirac's theory of first-class constraints was 
combined with the Feynman path integral method \cite{FADD,FAPO}. A general approach to quantization 
of first-class constraints systems was developed by Fradkin and his collaborators \cite{FRVI,FRAD}, it is reviewed in \cite{HENN}. 
An alternative symplectic scheme has been proposed by Faddeev and Jackiw \cite{FAJA,JACK}. 

Among physical systems under second-class constraints are anomalous gauge theories \cite{AFAD,JOOO,KOBA}, the nonlinear 
chiral model \cite{GALE}, many-body systems involving collective and independent-particle degrees of freedom \cite{MBODY1,MBODY2,MBODY3}. 
The Dirac's quantization method of such systems consists in constructing operators 
replicating the Dirac bracket for canonical variables and taking constraints to be operator equations. 
Batalin and Fradkin \cite{BATA} proposed a quantization scheme which converts constraints to first class 
by introducing new canonical variables. The problem reduces thereby to quantization of a first-class constraints system 
in an enlarged phase space. This method has found many useful applications.

The Hamilton-Jacobi procedure is also used for quantization of constrained systems \cite{HAJA,PIME,BALE}.


We show that each system of point particles with holonomic constraints has a hidden gauge symmetry which allows 
to perform its quantization without introducing new canonical variables. A mathematical pendulum on an $n-1$-dimensional 
sphere $S^{n-1}$ as an example of a mechanical system and the $O(n)$ nonlinear sigma model as an example of a field theory 
are quantized using the proposed method.

\vspace{2mm}

Let ${\mathcal L}$ be Lagrangian of point particles with holonomic constraints \cite{ARNO} 
\begin{equation}
{\mathcal L} = \frac{1}{2}{\dot \phi}^{\alpha}{\dot \phi}^{\alpha} - U + \lambda_{A}  \chi_{A} \label{LAGR}
\end{equation}
where $\lambda_{A}$ are Lagrange multipliers, $\chi_{A} = \chi_{A}(\phi^{\alpha}) = 0$, with $A = 1..m$, are constraints 
in the coordinate space $(\phi^{\alpha})$ with $\alpha = 1..n$, and $U = U( \phi^{\alpha})$ is a potential energy term. 

A more general class of systems with kinetic energy $ T = \frac{1}{2}\sum a_{i j}(q) {\dot q}^{i} {\dot q}^{j}$ 
can be converted to (\ref{LAGR}) by an isometric embedding 
into a euclidean space of a higher dimension \cite{NASH}.

The vectors $e^{\alpha}_{A}= \partial \chi_{A}/\partial \phi^{\alpha}$ constitute a local basis in a manifold 
orthogonal to the constraint surface $\chi_{A} = 0$. By linear transformations of $\chi_{A}$ with $\phi$-dependent 
coefficients, one can ensure the global orthogonality conditions $e^{\alpha}_{A}e^{\alpha}_{B} \approx \delta_{AB}$ \cite{PROOF}. 

A straightforward follow-up to the Dirac's scheme \cite{DIRA} leads to the following results: 
Let $\pi^{\alpha}$ and $\pi_{A}$ be canonical conjugated momenta corresponding to the canonical coordinates ${\phi}^{\alpha}$ and the 
Lagrange multipliers $\lambda_{A}$, respectively. The primary Hamiltonian is given by ${\mathcal H}_{p} = \frac{1}{2}\pi^{\alpha}\pi^{\alpha} 
+ U - \lambda_{A} \chi_{A}$. The primary constraint functions are $\Theta_{A} = \pi_{A}$. The canonical Hamiltonian is
${\mathcal H}_{c} = {\mathcal H}_{p} + u_{A}\pi_{A}$ where $u_{A}$ are undetermined functions of time. 
The holonomic constraints ${\chi}_{A} = 0$ appear again within the generalized Hamiltonian framework as secondary constraints: ${\chi}_{A} = \{ {\Theta}_{A}, {\mathcal H}_{c}\} \approx 0$.
The constraints of the next generation look like $\Omega_{A} = \{ {\chi}_{A}, {\mathcal H}_{c}\} = \pi^{\alpha} e^{\alpha}_{A}  \approx 0$,
while the third generation is 
$\nu_{A} = \{ {\Omega}_{A}, {\mathcal H}_{c}\} = \pi^{\alpha}\pi^{\beta} \partial^2 \chi_{A}/\partial \phi^{\alpha} \partial \phi^{\beta} 
- e^{\alpha}_{A} \partial U/\partial \phi^{\alpha} + e^{\alpha}_{A}e^{\alpha}_{B}\lambda_{B}  \approx 0$.
Equations ${\nu}_{A} = 0$ allow to fix $\lambda_{A}$ in terms of $\phi^{\alpha}$ and $\pi^{\alpha}$, no new constraints 
then appear. 

The phase space can be reduced by eliminating the $m$ canonical pairs $({\lambda}_{A},\pi_{A})$ from the Hamiltonian and 
the Poisson bracket. We solve equations ${\Theta}_{A} = 0$ with respect to $\pi_{A}$ and 
${\nu}_{A} = 0$ with respect to $\lambda_{A}$ and substitute solutions into ${\mathcal H}_{c}$. The result is 
\begin{equation}
{\mathcal H}_{c}^{\prime} = \frac{1}{2}\pi^{\alpha}\pi^{\alpha} + U
+ (\pi^{\alpha}\pi^{\beta} \frac{\partial^2 \chi_{A}}{\partial \phi^{\alpha} \partial \phi^{\beta}} - e^{\alpha}_{A} \frac{\partial U}{\partial \phi^{\alpha}}) 
(e^{\alpha}_{A}e^{\alpha}_{B})^{-1} \chi_{B}. 
\label{HCPRIME}
\end{equation} 

The remaining constraint functions ${\chi}_{A}$ and ${\Omega}_{A}$ satisfy
$\{\chi_{A}, \chi_{B} \} = 0$, $\{\chi_{A}, \Omega_{B} \} = e^{\alpha}_{A}e^{\alpha}_{B}$, and 
$\{ \Omega_{A}, \Omega_{B} \} = \partial e^{\beta}_{A} / \partial \phi^{\alpha} e^{\beta}_{B} \pi^{\alpha} 
                              - \partial e^{\beta}_{B} / \partial \phi^{\alpha} e^{\beta}_{A} \pi^{\alpha}$. 
These are second-class constraints. The ${\mathcal H}_{c}^{\prime}$ is first class. 

Two sets of the constraint functions are equivalent if they describe the same constraint surface. 
Two hamiltonian functions are equivalent i.e. provide the same hamiltonian dynamics if they and their first derivatives coincide on the constraint surface. 
E.g. the replacement $(e^{\alpha}_{A}e^{\alpha}_{B})^{-1} \rightarrow \delta_{AB}$ in Eq.(\ref{HCPRIME}) results to an equivalent 
Hamiltonian, since the variance is of the second order in the constraint functions. The replacements
$\pi^{\alpha} \rightarrow \pi^{\alpha}_s = \pi^{\alpha} - e^{\alpha}_{A} \Omega_{A}$ provide a possible modification also, etc.

The replacement $\Omega_{A} \rightarrow \Omega_{A}^{\prime} = \Omega_{A} - \partial e^{\beta}_{A} / \partial \phi^{\alpha} e^{\beta}_{B} \pi^{\alpha} \chi_{B}$
gives an equivalent set of the constraint functions ${\mathcal G}_{a} = (\chi_{A},\Omega_{A}^{\prime})$ with the Poisson bracket
\begin{equation}
\{ {\mathcal G}_{a},{\mathcal G}_{b} \} = {\mathcal I}_{ab} + {\mathcal C}_{ab\;}^{\;\;c} {\mathcal G}_{c}
\end{equation} 
where ${\mathcal I}_{AB} = {\mathcal I}_{m + A \; m + B} = 0$, 
${\mathcal I}_{A\;m + B} = -{\mathcal I}_{m + B \;A} = \delta_{AB}$, so that ${\mathcal I}_{ab} = - {\mathcal I}_{ba}$ 
and ${\mathcal I}_{ab}{\mathcal I}_{bc} = - {\delta}_{ac}$. 
In what follows, tensors with the upper and lower indices are related by ${\mathcal T}_{a} = {\mathcal I}_{ab}{\mathcal T}^{b}$.

The constraint and hamiltonian functions can be extrapolated from the constraint surface ${\mathcal G}_{a} = 0$ into the whole phase space in various ways. 
We wish to establish the normal form of the constraint and hamiltonian functions, ${\tilde {\mathcal G}}_{a}$ and ${\tilde {\mathcal H}}$, for which 
the relations $\{{\tilde {\mathcal G}}_{a}, {\tilde {\mathcal G}}_{b} \} = {\mathcal I}_{ab}$ and $\{{\tilde {\mathcal G}}_{a}, {\tilde {\mathcal H}}\} = 0$ 
are valid in the strong sense, at least locally near the constraint surface. 

Let us set ${\mathcal G}_{a}^{[1]} = {\mathcal G}_{a}$, ${\mathcal H}^{[1]} = {\mathcal H}^{\prime}_{c}$. The relations
\begin{eqnarray}
\{{\mathcal G}_{a}^{[k]}, {\mathcal G}_{b}^{[k]} \} &=& {\mathcal I}_{ab} + {\mathcal C}_{ab}^{\;\;c_{1}...c_{k}} {\mathcal G}_{c_{1}}^{[k]} ... {\mathcal G}_{c_{k}}^{[k]}, \label{GAGB} \\
\{{\mathcal G}_{a}^{[k]}, {\mathcal H}^{[k]} \} &=& {\mathcal R}_{a\;\;}^{\;b_{1}...b_{k}} {\mathcal G}_{b_{1}}^{[k]} ...{\mathcal G}_{b_{k}}^{[k]}, \label{GAH}
\end{eqnarray}
are valid for $k = 1$. If Eqs.(\ref{GAGB}) and (\ref{GAH}) are fulfilled for $k \geq 1$, the equivalent 
constraints and hamiltonian functions of order $k + 1$ can be constructed:

The Jacobi identities and symmetry of the structure functions ${\mathcal C}_{ab\;\;}^{\;\;c_{1}...c_{k}}$ and ${\mathcal R}_{a\;\;}^{\;b_{1}...b_{k}}$ 
with respect to permutations of the last $k$ indices give
\begin{eqnarray}
{\mathcal C}_{abc_{1}...c_{k}} + {\mathcal C}_{bc_{1}a...c_{k}} + {\mathcal C}_{c_{1}ab...c_{k}} &\approx& 0, \label{RRRN} \\
{\mathcal R}_{ab_{1}...b_{k}}  - {\mathcal R}_{b_{1}a...b_{k}} &\approx& 0.                    \label{ASYN}
\end{eqnarray} 
The next-order constraint and hamiltonian functions are given by 
\begin{eqnarray}
{\mathcal G}_{a}^{[k + 1]} &=& {\mathcal G}_{a}^{[k]} + 
\frac{1}{k + 2} {\mathcal C}_{a}^{\;c_{0}c_{1}...c_{k}} {\mathcal G}_{c_{0}}^{[k]} {\mathcal G}_{c_{1}}^{[k]} ... {\mathcal G}_{c_{k}}^{[k]}, \label{GPLUS} \\
{\mathcal H}^{[k + 1]} &=& {\mathcal H}^{[k]} - \frac{1}{k + 1} {\mathcal R}^{b_{0}b_{1}...b_{k}} {\mathcal G}_{b_{0}}^{[k]} {\mathcal G}_{b_{1}}^{[k]} ... {\mathcal G}_{b_{k}}^{[k]}. \label{HPLUS}
\end{eqnarray}
They define the same constraint surface and the same hamiltonian flow on the constraint surface as one step before.
Using Eqs.(\ref{GAGB}) and (\ref{GAH}), one can calculate the next-order structure functions 
${\mathcal C}_{ab\;\;}^{\;\;c_{1}...c_{k + 1}}$ and ${\mathcal R}_{a\;\;}^{\;b_{1}...b_{k + 1}}$ and repeat the procedure. 
If structure functions vanish, we shift the $k$ by one unit and check Eqs.(\ref{GAGB}) and (\ref{GAH}) again.
At each step, the Poisson bracket relations get closer to the normal form. 
In the limit $k \rightarrow + \infty$, we obtain
$\{{\tilde {\mathcal G}}_{a}, {\tilde {\mathcal G}}_{b} \} = {\mathcal I}_{ab}$ and 
$\{{\tilde {\mathcal G}}_{a}, {\tilde {\mathcal H}}\} = 0$. A similar theorem on the normal form exists for systems under first-class constraints \cite{HENN}.

The $\chi_{A}$ depend on the $\phi^{\alpha}$ only, $\Omega_{A}^{\prime} \sim \pi^{\alpha}$, and ${\mathcal H}_{c}^{\prime} = T + V$ where
$T \sim \pi^{\alpha}\pi^{\beta}$ and $V \sim 1$. These 
properties propagate throughout the above algorithm, being thus inherent for ${\tilde {\mathcal G}}_{a}$ and ${\tilde {\mathcal H}}$: 

Indeed, let $I,J,K..=1..m$ and $Q,R,S..=m+1..2m$. 
Let at an $k$-th step ${\mathcal G}_{I}$ be a function of $\phi^{\alpha}$ and ${\mathcal G}_{Q}$ be linear in $\pi^{\alpha}$. 
By inspection of Eq.(\ref{GAGB}) one can conclude that 
${\mathcal C}_{I}^{\;Qc_{1}...c_{k}} = 0$ for all $c_{i}$,
${\mathcal C}_{I}^{\;Jc_{1}...c_{k}} \ne 0$ only if $c_{i} = K,L,M..$. 
The coefficient ${\mathcal C}_{Q}^{\;Ic_{1}...c_{k}}$ can be expanded into zero- and first-order terms in $\pi^{\alpha}$.
The zeroth term does not vanish if $c_{i} = J,K,L..$ at $i = 1..k$ except for $i=l$ with $c_{l} = R,S,T..$.
The first-order term does not vanish if $c_{i} = J,K,L..$ at $i = 1..k$. 
Eq.(\ref{GPLUS}) implies that ${\mathcal G}_{I} \sim 1$ and ${\mathcal G}_{Q} \sim \pi^{\alpha}$ at step $k+1$.

Eq.(\ref{GAH}) is valid at $k = 1$ separately for $T$ and $V$. 
Let ${ T}^{[k]} \sim \pi^{\alpha}\pi^{\beta}$. By inspection of Eq.(\ref{GAGB}) one finds that
structure functions of the kinetic energy $T$ obey 
${\mathcal R}_{I}^{\;J...K} \sim \pi^{\alpha}$,
${\mathcal R}_{I}^{\;QJ...K} \sim 1$,
${\mathcal R}_{I}^{\;QRJ...K} = 0$,
${\mathcal R}_{Q}^{\;J...K} \sim \pi^{\alpha}\pi^{\beta}$,
${\mathcal R}_{Q}^{\;RJ...K} \sim \pi^{\alpha}$,
${\mathcal R}_{Q}^{\;RSJ...K} \sim 1$, and
${\mathcal R}_{Q}^{\;RSTJ...K} = 0$. 
Eq.(\ref{HPLUS}) implies that the next-order kinetic part of the Hamiltonian remains quadratic with respect to $\pi^{\alpha}$.
Let further ${ V}^{[n]} \sim 1$. Eq.(\ref{GAGB}) tells that structure functions of the potential energy $V$ satisfy
${\mathcal R}_{Q}^{\;IJ...K} \sim 1$, the other functions vanish. 
In virtue of Eq.(\ref{HPLUS}), the next-order $V$ is again a potential.

The above algorithm assures 
${\mathcal G}_{a}         \approx {\tilde {\mathcal G}}_{a}$, 
${\mathcal H}_{c}^{\prime} \approx {\tilde {\mathcal H}}    $, 
$\{ f, {\mathcal G}_{a} \} \approx \{ f, {\tilde {\mathcal G}}_{a}\}$, and 
$\{ f, {\mathcal H}_{c}^{\prime} \} \approx \{ f, {\tilde {\mathcal H}} \}$ 
for arbitrary functions $f$. 
In the basis ${\tilde {\mathcal G}}_{a} = ({\tilde \chi_{A}},{\tilde \Omega_{A}})$, 
$E^{\alpha}_{A} = \partial {\tilde \chi_{A}}/ \partial \phi^{\alpha}$. Since ${\tilde \Omega_{A}}$ is linear in the momenta, 
it can be written as ${\tilde \Omega_{A}} = \pi ^{\alpha} F^{\alpha}_{A}$. The normal form 
of the constraints implies $E^{\alpha}_{A} F^{\alpha}_{B} = \delta_{AB}$ and 
\begin{equation}
F^{\beta}_{B}\partial F^{\alpha}_{A}/\partial \phi^{\beta} = F^{\beta}_{A}\partial F^{\alpha}_{B}/\partial \phi^{\beta}.
\label{NORM}
\end{equation} 
In addition, we have $e^{\alpha}_{A} \approx E^{\alpha}_{A} \approx F^{\alpha}_{A}$.

The constraint functions ${\tilde {\mathcal G}}_{a}$ generate transformations of the canonical variables
\begin{eqnarray}
\delta \phi^{\alpha} &=& \{ \phi^{\alpha} , {\tilde \chi_{A}} \} \theta_{A} = 0,  \label{11} \\
\delta \pi ^{\alpha} &=& \{ \pi ^{\alpha} , {\tilde \chi_{A}} \} \theta_{A} = - E^{\alpha}_{A}\theta_{A}, \label{12} \\
\Delta \phi^{\alpha} &=& \{ \phi^{\alpha} , {\tilde \Omega_{A}} \} \eta_{A} = F^{\alpha}_{A}\eta_{A}, \label{21}\\
\Delta \pi ^{\alpha} &=& \{ \pi ^{\alpha} , {\tilde \Omega_{A}} \} \eta_{A} = - \pi^{\beta} \partial F^{\beta}_{A}/\partial \phi^{\alpha} \eta_{A}, \label{22}
\end{eqnarray}
with respect to which ${\tilde {\mathcal H}}$ is invariant, with $\theta_{A}$ and $\eta_{A}$ being infinitesimal parameters.

Lagrangian ${\tilde {\mathcal L}}(\phi^{\alpha},\pi^{\alpha},{\dot \phi}^{\alpha}) = \pi^{\alpha} {\dot \phi}^{\alpha} - {\tilde {\mathcal H}}$ calculated 
for ${\dot \phi}^{\alpha} = \{ {\phi}^{\alpha}, {\tilde {\mathcal H}}\}$ is invariant under (\ref{11}) - (\ref{22}) up to a total differential 
of some function. Suppose we have solved
the ${\dot \phi}^{\alpha} = \{ {\phi}^{\alpha}, {\tilde {\mathcal H}}\}$ for the ${\pi}^{\alpha}$ and got 
${\tilde {\mathcal L}}(\phi^{\alpha},{\dot \phi}^{\alpha}) = {\tilde {\mathcal L}}(\phi^{\alpha},\pi^{\alpha}(\phi^{\beta},{\dot \phi}^{\beta}),{\dot \phi}^{\alpha})$.
The transformation properties the ${\dot \phi}^{\alpha}$ are as follows
\begin{eqnarray}
\delta {\dot \phi}^{\alpha} &=& \{ \{ \phi^{\alpha}, {\tilde {\mathcal H}} \}, {\tilde \chi_{A}} \} \theta_{A} = 0, \label{31} \\
\Delta {\dot \phi}^{\alpha} &=& \{ \{ \phi^{\alpha}, {\tilde {\mathcal H}} \}, {\tilde \Omega_{A}} \} \eta_{A}      
= \partial F^{\alpha}_{A}/\partial \phi^{\beta} {\dot \phi}^{\beta} \eta_{A}.  \label{32}
\end{eqnarray}
Lagrangian ${\tilde {\mathcal L}}(\phi^{\alpha},{\dot \phi}^{\alpha})$ as a function in the tangent bundle $(\phi^{\alpha},{\dot \phi}^{\alpha})$
is invariant with respect to the $m$-parameter family of the global transformations 
(\ref{21}) and (\ref{32}). 
Equations $\{ {\tilde \Omega}_{A},{\tilde {\mathcal H}} \} = 0$ are equivalent to the following equations
\begin{equation}
\frac{ \partial {\tilde {\mathcal L}}}{\partial {\phi}^{\alpha}} F^{\alpha}_{A} + 
\frac{ \partial {\tilde {\mathcal L}}}{\partial {\dot \phi}^{\alpha}} 
\frac{ \partial F^{\alpha}_{A}}{\partial {\phi}^{\beta}} {\dot \phi}^{\beta} = 0 
\end{equation}
which indicate explicitly \cite{COMM} the invariance of ${\tilde {\mathcal L}}$ against the 
transformations (\ref{21}) and (\ref{32}). 

The constraint functions ${\tilde \chi_{A}}$ do not induce transformations in the tangent bundle $(\phi^{\alpha},{\dot \phi}^{\alpha})$. 
This is a specific feature of holonomic systems whose constraints do not depend on velocities. 
Equations $\{ {\tilde \chi_{A}}, {\tilde {\mathcal H}} \} = 0$ imply, however, $E^{\alpha}_{A} {\dot \phi}^{\alpha} = 0$. 
Consequently, ${\tilde {\mathcal L}}(\phi^{\alpha},{\dot \phi}^{\alpha})$ is defined on the constraint surface 
$E^{\alpha}_{A} {\dot \phi}^{\alpha} = 0$. It can be extrapolated from the constraint surface according to
${\tilde {\mathcal L}}^{\prime}(\phi^{\alpha},{\dot \phi}^{\alpha}) = {\tilde {\mathcal L}}(\phi^{\alpha},\Delta^{\alpha \beta}{\dot \phi}^{\beta})$
where $\Delta^{\alpha \beta} = \delta^{\alpha \beta} - F^{\alpha}_{A} E^{\beta}_{A}$, $\Delta^{\alpha \beta}$ acts as a projection 
operator. ${\tilde {\mathcal L}}^{\prime}$
is invariant under an $m$-parameter family of transformations
\begin{eqnarray}
\delta^{\prime} \phi^{\alpha} &=& 0, \label{41} \\
\delta^{\prime} {\dot \phi}^{\alpha} &=& F^{\alpha}_{A} \vartheta_{A}  \label{42}
\end{eqnarray}
where $\vartheta_{A}$ are infinitesimal parameters.
Equations ${\tilde \Omega}_{A} = \pi^{\alpha}F^{\alpha}_{A} = 0$ are equivalent to the following equations
\begin{equation}
\frac{ \partial {\tilde {\mathcal L}}}{\partial {\dot \phi}^{\alpha}} F^{\alpha}_{A} = 0 \label{DOT}
\end{equation} 
which show the invariance 
of ${\tilde {\mathcal L}}$ under transformations (\ref{41}) and (\ref{42}). Lagrangian ${\tilde {\mathcal L}}$, as a 
Legendre transform of ${\tilde {\mathcal H}}$, depends thereby on the tangent velocities $\Delta^{\alpha \beta} {\dot \phi}^{\beta}$ automatically, i.e.,
${\tilde {\mathcal L}}^{\prime}(\phi^{\alpha},{\dot \phi}^{\alpha}) = {\tilde {\mathcal L}}(\phi^{\alpha},{\dot \phi}^{\alpha})$.

The invariance of Lagrangian ${\tilde {\mathcal L}}$ under the $2m$-parameter family of the global transformations (\ref{21}) and (\ref{32}), 
(\ref{41}) and (\ref{42}) 
provides the invariance under an $m$-parameter family of local transformations of trajectories $\phi^{\alpha}(t)$. Let 
$\eta_{A}$ be time dependent functions and $\vartheta_{A} = {\dot \eta_{A}}$. The transformation 
law in the tangent bundle $(\phi^{\alpha},{\dot \phi}^{\alpha})$ reads
\begin{eqnarray}
\delta^{\prime}       \phi ^{\alpha} + \Delta       \phi ^{\alpha} &=& F^{\alpha}_{A}       \eta_{A}, \label{NDOT} \\
\delta^{\prime} {\dot \phi}^{\alpha} + \Delta {\dot \phi}^{\alpha} &=& F^{\alpha}_{A} {\dot \eta_{A}} + \frac{\partial F^{\alpha}_{A}}{\partial \phi^{\beta}} {\dot \phi}^{\beta} \eta_{A}
                                                          = \frac{d}{dt} \left( F^{\alpha}_{A}\eta_{A} \right ). \label{WDOT}
\end{eqnarray}
The last equation suggests that this transformation acts on the trajectories $\phi^{\alpha}(t)$. Lagrangian
${\tilde {\mathcal L}}$ is therefore invariant with respect to the local transformations. The variations of $\phi^{\alpha}(t)$
along the vectors $F^{\alpha}_{A}$ correspond to the $m$ gauge degrees 
of freedom. The constraints $\chi_{A} = 0$ play a role of the gauge-fixing conditions. The gauge-invariant observables are projections 
of trajectories $\phi^{\alpha}(t)$ onto the constraint surface $\chi_{A} = 0$. Eq.(\ref{NORM}) ensures that the symmetry group is abelian.

If ${\tilde {\mathcal L}}$ would be chosen as the starting Lagrangian to pass over to the Hamiltonian framework, one could get 
${\tilde {\mathcal H}}$ as primary Hamiltonian and ${\tilde \Omega}_{A} = \pi^{\alpha}F^{\alpha}_{A} = 0$ as primary first-class 
constraints by virtue of Eq.(\ref{DOT}). The constraints ${\tilde \chi_{A}} = 0$ 
can be selected as a set of the admissible gauge-fixing conditions. Starting from Lagrangian (\ref{LAGR}), we arrive at the set of second-class 
constraints ${\tilde \chi_{A}} = 0$ and ${\tilde \Omega_{A}} = 0$. Starting from ${\tilde {\mathcal L}}$, we obtain the same set as 
gauge-fixing conditions ${\tilde \chi_{A}} = 0$ and first-class constraints ${\tilde \Omega_{A}} = 0$ associated with the gauge symmetry. It shows that 
classification of constraints in the holonomic systems is a matter of convention.

{\it 
As Lagrangian (\ref{LAGR}) is equivalent to its gauged counterpart ${\tilde {\mathcal L}}(\phi^{\alpha},{\dot \phi}^{\alpha})$ under the
gauge-fixing conditions $\chi_{A} = 0$, one can apply one of the quantization schemes specific for gauge theories. 
}

The system is quantized by the algebra mapping $(\phi^{\alpha}, \pi^{\alpha}) \to ({\hat \phi}^{\alpha},{\hat \pi}^{\alpha})$ 
and $\{,\}\to -i[,]$. Consequently, to any symmetrized function in the phase space variables one may associate an 
operator function. The function is symmetrized in such a way that quantal image is a hermitian operator.

The wave functions obey the Dirac's supplementary condition
\begin{equation}
{\hat {\tilde \Omega}}_{A} \Psi = 0. \label{DIRA}
\end{equation}
In the quasiclassical limit, gradients of the wave functions along the vectors $F^{\alpha}_{A}$ vanish. 

The path integral for the evolution operator becomes
\begin{eqnarray}
&&Z =  \label{PARQ} \\
&&\int{ \prod _{} d^n \phi d^n \pi \prod_{A} \delta({\tilde \chi}_{A}) \delta({\tilde \Omega}_{A}) }
\exp \left \{ i \int{dt(\pi^{\alpha} {\dot \phi}^{\alpha} - {\tilde {\mathcal H}})} \right \}. \nonumber
\end{eqnarray}
Eqs.(\ref{DIRA}) and (\ref{PARQ}) solve the quantization problem for systems of point particles with holonomic constraints. 

Given the initial state wave function $\Psi(0)$ satisfying (\ref{DIRA}), the final wave function $\Psi(t)$ can be 
found applying the evolution operator (\ref{PARQ}) on $\Psi(0)$. Since ${\hat {\tilde \Omega}}_{A}$ commute with the Hamiltonian, the final 
state wave function obey Eqs.(\ref{DIRA}).

It is desirable that the quantization procedure does not destroy the classical symmetries which results in having the supplementary 
condition (\ref{DIRA}) satisfied by the state $\Psi(t)$ for any value of $t$. This feature can apparently be violated either due to 
complex terms entering ${\tilde {\mathcal H}}$ or by approximations adopted for treating the operator eigenvalues. The supplementary 
condition (\ref{DIRA}) assures, however, that the symmetry is preserved in average. As we shall see latter on, the quantization procedure 
preserves the gauge symmetries explicitly in the applications considered.

The supplementary condition (\ref{DIRA}) does not comprise the constraints ${\tilde \chi}_{A} = 0$.
In gauge theories, the evolution operator is independent of the gauge-fixing conditions \cite{FADD}. 
We can insert $\det \{ {\tilde \chi}_{A}, {\tilde \Omega}_{B}\} = 1$ into the integrand of Eq.(\ref{PARQ}) 
to bring the measure into the form identical to the one prescribed by gauge theories. 
In the path integral, the ${\tilde \chi}_{A}$ can then be replaced with arbitrary functions.

{\it
It is remarkable that physical observables are determined only by half of the number of second-class constraints.
}

\vspace{2mm}

Let us illustrate the method proposed with quantization of a mathematical pendulum on an $n-1$-dimensional sphere $S^{n-1}$.

Lagrangian is given by Eq.(\ref{LAGR}) with $m=1$ and $\chi = \ln \phi$. Using the Dirac scheme, we obtain the primary Hamiltonian 
${\mathcal H}_{p} = \frac{1}{2}\pi^2 - \lambda \ln \phi$ and the primary constraint ${\mathcal G}_{0} = \pi_{\lambda} \approx 0$. The canonical 
Hamiltonian becomes ${\mathcal H}_{c} = {\mathcal H}_{p} + u\pi_{\lambda}$ where $u$ is an undetermined function of time. The secondary constraints 
${\mathcal G}_{a+1} = \{ {\mathcal G}_{a}, {\mathcal H}_{c}\}$ can be found: 
${\mathcal G}_{1} = \ln \phi$, ${\mathcal G}_{2} = \phi \pi/\phi^2$, ${\mathcal G}_{3} = \pi^2/\phi^2 - 2(\phi\pi)^2/ \phi^4 + \lambda/\phi^2$. The last 
constraint ${\mathcal G}_{3} = 0$ allows to fix $\lambda$ in terms of $\phi^{\alpha}$ and $\pi^{\alpha}$, no new constraints then appear. 
The canonical pair $({\lambda},\pi_{\lambda})$ is eliminated by solving equations ${\mathcal G}_{0} = 0$ and ${\mathcal G}_{3} = 0$. The result 
is ${\mathcal H}_{c}^{\prime} = \frac{1}{2}\pi^2 + (\pi^2 - 2(\phi\pi)^2/ \phi^2) \ln \phi$. There remain two constraint 
functions ${\mathcal G}_{1}$ and ${\mathcal G}_{2}$, such that $\{ {\mathcal G}_{1} ,{\mathcal G}_{2}\} = 1/\phi^2$. These are second-class 
constraints. The Hamiltonian ${\mathcal H}_{c}^{\prime}$ is first class.
 
In order to split the constraints into a gauge-fixing condition and a gauge generator, we should construct 
${\tilde {\mathcal G}}_{a}$ and ${\tilde {\mathcal H}}$ starting from ${\mathcal G}_{a}$ and ${\mathcal H}_{c}^{\prime}$. 
The same constraint surface is described by the functions ${\tilde \chi} = \ln \phi$ 
and ${\tilde \Omega} = \phi \pi$ satisfying $\{ {\tilde \chi} ,{\tilde \Omega} \} = 1$, so one can set ${\tilde {\mathcal G}}_{1} = \ln \phi$ and 
${\tilde {\mathcal G}}_{2} = \phi \pi$. 

The Hamiltonian ${\tilde {\mathcal H}}$ can be found to be 
\begin{equation}
{\tilde {\mathcal H}} = \frac{1}{2} \phi^2 \Delta^{\alpha \beta} \pi^{\alpha} \pi^{\beta}
\end{equation}
where $\Delta^{\alpha \beta} = \delta^{\alpha \beta} - \phi^{\alpha} \phi^{\beta} /\phi^2$. For $n=3$, 
${\tilde {\mathcal H}} = \frac{1}{2}L^{\alpha}L^{\alpha}$ where $L^{\alpha}$ is the orbital momentum. The difference 
${\tilde {\mathcal H}} - {\mathcal H}_{c}^{\prime}$ is of the second order in the constraint functions. This is
sufficient to have identical hamiltonian flows on the constraint surface.

${\tilde \Omega}$ is stable with respect to time evolution $\{ {\tilde \Omega}, {\tilde {\mathcal H}} \} = 0$. The relations
$\Delta \phi^{\alpha} = \{ \phi^{\alpha} , {\tilde \Omega} \} =   \phi^{\alpha}$ and 
$\Delta \pi ^{\alpha} = \{ \pi ^{\alpha} , {\tilde \Omega} \} = - \pi ^{\alpha}$
show that ${\tilde \Omega}$ generates dilatations of $\phi^{\alpha}$ and contractions of $\pi^{\alpha}$. 
The ${\tilde \chi}$ is identically in involution with the Hamiltonian $\{ {\tilde \chi}, {\tilde {\mathcal H}} \} = 0$.
The Poisson bracket relations $\{ \phi^{\alpha} , {\tilde \chi} \} =   0$ and $\{ \pi ^{\alpha} , {\tilde \chi} \} = - \phi ^{\alpha}$
define shifts of the longitudinal component of the canonical momenta.

The gauged Lagrangian is given by
\begin{equation} \label{GLAG}
{\tilde {\mathcal L}} = \frac{1}{2} \Delta^{\alpha \beta}(\phi){{\dot \phi}^{\alpha} {\dot \phi}^{\beta}}/{\phi^{2}}. 
\end{equation}
It is invariant under the local dilatations of $\phi^{\alpha}$. In the context of the dynamics defined by ${\tilde {\mathcal L}}$
with no constraints imposed, $\phi$ is an arbitrary function of time. It can always be selected to fulfill the constraint 
$\chi = 0$ or some other admissible constraint. 
The Euler-Lagrange equations have the form
\begin{equation}
\Delta^{\alpha \beta} \frac{d^2}{dt^2} (\phi^{\beta}/{\phi} )  = 0. \label{EM}
\end{equation}
The point particle moves thereby without tangent acceleration. Eqs.(\ref{EM}) are not affected by constraints. 

${\tilde {\mathcal L}}$ depends on the spherical coordinates $\phi^{\alpha}/{\phi}$ 
which lie on an $n-1$-dimensional sphere of a unit radius. Eqs.(\ref{EM}) for unconditional extremals 
in the coordinate space $\phi^{\alpha}$ coincide therefore with the D'Alembert-Lagrange equations for 
extremals in the spherical coordinates space, considered as independent Lagrange variables,
under the condition that they lie on an $n-1$-dimensional sphere of a unit radius.

The constraint $ {\tilde \chi} = 0$ can therefore
be treated as a gauge-fixing condition, the function $\phi$ as a gauge degree of freedom, the ratios $\phi^{\alpha}/{\phi}$ as 
gauge invariant observables. The equations of motion (\ref{EM}) are formulated in terms of the gauge invariant observables.
The extremals of the action functionals  $\int{{\mathcal L}dt}$ and 
$\int{{\tilde {\mathcal L}}dt}$ under the constraint $ {\tilde \chi} = 0$ coincide.

The quantum hermitian Hamiltonian has the form 
${\hat {\tilde {\mathcal H}}} = \frac{1}{2} \phi \Delta^{\alpha \beta} {\hat \pi}^{\beta} \phi \Delta^{\alpha \gamma} {\hat \pi}^{\gamma}$. 
The vector $i\phi \Delta^{\alpha \beta} {\hat \pi}^{\beta}$ gives the angular part of the gradient operator. 
Although it is not conspicuous, ${\hat {\tilde {\mathcal H}}}$ does not depend on the radial coordinate $\phi$.
The constraint operator can be defined as ${\hat {\tilde \Omega}} = (\phi^{\alpha}{\hat \pi}^{\alpha} + {\hat \pi}^{\alpha}\phi^{\alpha})/2$. 
It acts only on the radial component of $\phi^{\alpha}$, so the relation $[{\hat {\tilde \Omega}},{\hat {\tilde {\mathcal H}}}] = 0$ 
holds.
The supplementary condition (\ref{DIRA}) implies $\Psi = \phi^{-n/2}\Psi_{1}(\phi^{\alpha}/\phi)$. The important information about the system
is contained in $\Psi_{1}(\phi^{\alpha}/\phi)$ only. 

The integral over the canonical momenta Eq.(\ref{PARQ}) can be evaluated explicitly to give
\begin{equation}
Z = \int{ \prod _{} \sqrt{\left ({\partial  {\tilde \chi}}/{\partial \phi^{\alpha}} \right )^2}\delta( {\tilde \chi}) d^n\phi 
\exp \left \{ i \int{dt  {\tilde {\mathcal L}} } \right \} }. \label{MEAS}
\end{equation}
Lagrange's measure $\sqrt{\left ({\partial  {\tilde \chi}}/{\partial \phi^{\alpha}} \right )^2}\delta( {\tilde \chi}) d^n\phi$ coincides with the
volume element of the $S^{n-1}$ sphere. It can be rewritten as an invariant volume of the configuration space e.g. in terms of 
the angular variables $(\varphi_{1}...\varphi_{n-1})$ with the help of the induced metric tensor. 
Lagrange's measure is invariant under the group $O(n)$. It remains the same for all functions $ {\tilde \chi}$ vanishing for $\phi = 1$.

\vspace{2mm}

The $O(n)$ nonlinear sigma model represents the field theory analog of the spherical $n-1$-dimensional pendulum. The $n=4$ case 
corresponds to the chiral nonlinear sigma model which is invariant under the $SU(2)_L\otimes SU(2)_R \sim SO(4)$ group. 
Its quantization does not involve essentially new ingredients. The coordinates $\phi^{\alpha}$ become functions of $x_{\mu}=(t,x,y,z)$,
the time derivatives ${\dot \phi}^{\alpha}$ in Eqs.(\ref{LAGR}) and (\ref{GLAG}) should be replaced with $\partial_{\mu} \phi^{\alpha}$. The wave 
functionals depend on the spherical coordinates $\phi^{\alpha}(x)/\phi(x)$, the path integral has the form of Eq.(\ref{MEAS}) with $dt$
replaced by $d^4x$.

{\it The quantized gauged pendulum and $O(n)$ nonlinear sigma model preserve the hidden gauge symmetries. 
Consequently, the wave functions $\Psi$ satisfy the Dirac's supplementary condition at any time, i.e., the systems 
evolve always on the constraint surfaces.}

\vspace{2mm}

We showed that systems of point particles with holonomic constraints have hidden abelian gauge symmetries. 
Their physical origin is the following: 
We allow virtual displacements of the particles from the constraint surface and treat such displacements as unphysical gauge degrees of freedom. 
The holonomic constraints turn thereby into gauge-fixing conditions. 
The physical variables are fully specified by projecting the particle coordinates onto the constraint surface. 
In the considered examples, the projected coordinates are the spherical coordinates $\phi^{\alpha}/\phi$.
The existence of the equivalent gauge models allows us to quantize the original holonomic systems in their own phase space using
standard methods of quantization known from the gauge theories. 

After submitting this work, our attention has been called to papers \cite{ABDA01,BANE94,MITRA1,VYTHE1,VYTHE2,VYTHE3}.
In Ref. \cite{ABDA01}, a mathematical pendulum is quantized using the Dirac approach. The $O(n)$ non-linear sigma 
model is discussed along the line \cite{BATA} in Ref. \cite{BANE94}. In Ref. \cite{MITRA1}, hidden gauge symmetries 
of second class constraints systems are discussed and a quantization method in the original phase space and an extended 
configuration space is proposed. The holonomic systems, as we showed, admit the quantization in the original phase and 
configuration spaces. The methods \cite{BATA,MITRA1} are applyed for quantization of various physical systems 
and compared in Refs. \cite{VYTHE1,VYTHE2,VYTHE3}. 

\vspace{2mm}

M.I.K. and A.A.R. wish to acknowledge kind hospitality at the University of Tuebingen. This work is supported by the Helmholtz Gemeinschaft grant
for GSI (Darmstadt), DFG grant No. 436 RUS 113/721/0-1 and RFBR grant No. 03-02-04004.




\begin{thebibliography}{00}
\bibitem{PROOF} Let 
$\chi_{1}^{\prime} = \chi_{1}$, $e^{\prime \alpha}_{1} = e^{\alpha}_{1}$, and
$\chi_{A + 1}^{\prime} = - \sum_{B=1}^{A} ({e^{\prime \alpha}_{B} e^{\alpha}_{A + 1}}/{e^{\prime \beta}_{B}e^{\prime \beta}_{B}}) \chi_{B}^{\prime} + \chi_{A + 1},$
$e^{\prime \alpha}_{A + 1} = \partial \chi_{A + 1}^{\prime}/\partial \phi^{\alpha}.$
The constraint functions $\chi_{A}^{\prime}$ defined above recursively describe the same constraint surface $\chi_{A} = 0$, while the vectors $e^{\prime \alpha}_{A}$ are orthogonal 
in the weak sense. The functions $\chi_{A}^{\prime \prime} = \chi_{A}^{\prime}/\sqrt{e^{\prime \alpha}_{B}e^{\prime \alpha}_{B}}$ define further a set of the
normalized orthogonal vectors $e^{\prime \prime \alpha}_{A}$.
\bibitem{COMM} Indeed, $\partial {\tilde {\mathcal L}}/\partial {\phi}^{\alpha} = 
\frac{d}{dt}\partial {\tilde {\mathcal L}}/\partial {\dot \phi}^{\alpha} = \frac{d}{dt} \pi^{\alpha} = \{ \pi^{\alpha}, {\tilde {\mathcal H}}\}$. The second term
can be rewritten as 
$\pi^{\alpha} \partial F^{\alpha}_{A}/\partial {\phi}^{\beta} {\dot \phi}^{\beta} = \pi^{\alpha} \{ F^{\alpha}_{A}, {\tilde {\mathcal H}}\}$. The sum
$ F^{\alpha}_{A} \{ \pi^{\alpha}, {\tilde {\mathcal H}}\} + \pi^{\alpha} \{ F^{\alpha}_{A}, {\tilde {\mathcal H}}\}$ gives 
$\{ {\tilde \Omega}_{A},{\tilde {\mathcal H}} \}$.
\end{thebibliography}
\end{document}